\documentclass[english,reprint,showpacs,preprintnumbers,showkeys,amsmath,amssymb,aps,floatfix,prl]{revtex4-1}
\usepackage[latin9]{inputenc}
\usepackage{color}
\usepackage{mathrsfs}
\usepackage{amsmath}
\usepackage{graphicx}

\makeatletter

\@ifundefined{textcolor}{}
{%
\definecolor{BLACK}{gray}{0}
\definecolor{WHITE}{gray}{1}
\definecolor{RED}{rgb}{1,0,0}
\definecolor{GREEN}{rgb}{0,1,0}
\definecolor{BLUE}{rgb}{0,0,1}
\definecolor{CYAN}{cmyk}{1,0,0,0}
\definecolor{MAGENTA}{cmyk}{0,1,0,0}
\definecolor{YELLOW}{cmyk}{0,0,1,0}
}

\usepackage{babel}

\@ifundefined{showcaptionsetup}{}{%
 \PassOptionsToPackage{caption=false}{subfig}}
\usepackage{subfig}
\makeatother

\usepackage{babel}
\begin{document}

\title{Unwinding the hairball graph: pruning algorithms for \\
 weighted complex networks}

\author{Navid Dianati}

\thanks{n.dianatimaleki@neu.edu}

\affiliation{The Lazer Lab, Northeastern University, Boston Massachusetts.}

\affiliation{Institute for Quantitative Social Sciences, Harvard University, Cambridge
Massachusetts.}
\begin{abstract}
Empirical networks of weighted dyadic relations often contain ``noisy''
edges that alter the global characteristics of the network and obfuscate
the most important structures therein. Graph pruning is the process
of identifying the most significant edges according to a generative
null model, and extracting the subgraph consisting of those edges.
\textcolor{black}{Here, we focus on integer-weighted graphs commonly
arising when weights count the occurrences of an ``event'' relating
the nodes.} We introduce a simple and intuitive null model related
to the configuration model of network generation, and derive two significance
filters from it: the Marginal Likelihood Filter (MLF) and the Global
Likelihood Filter (GLF). The former is a fast algorithm assigning
a significance score to each edge based on the marginal distribution
of edge weights whereas the latter is an ensemble approach which takes
into account the correlations among edges. We apply these filters
to the network of air traffic volume between US airports and recover
a geographically faithful representation of the graph. Furthermore,
compared with thresholding based on edge weight, we show that our
filters extract a larger and significantly sparser giant component. 
\end{abstract}

\keywords{weighted networks, statistical significance, ERGM, exponential random
graph model}

\maketitle

\section{Introduction}

Graphs or networks are widely used as representations of the structure
and dynamics of complex systems \cite{strogatz_exploring_2001,albert_statistical_2002,barthelemy_spatial_2011,jin_structure_2001,newman_structure_2003}.
Too often in practice, networks of observed dyadic relationships are
too dense to be of immediate use: the topology of the network is dominated
by an abundance of ``noisy'' edges that must somehow be removed
before the most significant structures are revealed. This process---which
we refer to as \textit{pruning}---is particularly useful in visualizing
the so-called ``hairball'' networks, and can conceivably \textcolor{black}{enhance
the efficacy of community detection methods by serving as a preconditioner.}

\textcolor{black}{We distinguish between the problem of pruning discussed
here on the one hand, and the problem of }\textit{\textcolor{black}{sparsification
}}\textcolor{black}{on the other. Sparsification is the problem of
approximating a network using a subgraph with fewer edges such that
some property of the graph is preserved within a desired tolerance.
The goal of sparsification is typically to compute network characteristics
of the original graph, only at a lower computational cost. Therefore,
one must aim to minimally alter the character of the network in the
process. For instance, when faced with a dense similarity matrix derived
from a large number of data points, it is desirable to work instead
with a sparse subgraph with the same community structure as the full
graph. For such applications, one may use }\textit{\textcolor{black}{sparsifiers
}}\textcolor{black}{using random spanning trees \cite{goyal_expanders_2009,kelner_faster_2009,wilson_generating_1996},
or others that explicitly approximate the spectral properties of the
graph Laplacian \cite{spielman_spectral_2008}.}

\textcolor{black}{The problem of }\textit{\textcolor{black}{pruning
}}\textcolor{black}{on the other hand, involves the removal of a possibly
large number of spurious edges that are believed to obfuscate an unknown
core that contains the most important structures. It is therefore
implied that the coveted core is }\textit{\textcolor{black}{different
}}\textcolor{black}{from the observed, noisy graph. The properties
of the core such as its community structure are not known }\textit{\textcolor{black}{a
priori}}\textcolor{black}{, and thus, it is not clear which graph
properties if any should be preserved in the process. In fact, the
goal should arguably be to }\textit{\textcolor{black}{alter }}\textcolor{black}{important
features of the graph until the properties of the hidden core are
revealed.}

Graph pruning is most commonly done by thresholding based on edge
weights. This approach equates significance with edge weight, and
fails to take into account the relationship between the edge, its
incident vertices and their other edges. Therefore, thresholding based
on weight systematically discounts low-degree vertices and structures
they represent. In order to address this issue, alternative methods
have been proposed such as the filters of \cite{serrano_extracting_2009}
and \cite{radicchi_information_2011}. These methods consist of assigning
a \textit{p}-value to each edge based on a null model of edge weight
distribution, and subsequently filtering out all but those edges least
likely to have occurred due to pure chance, namely those with the
smallest \textit{p}-values. \textcolor{black}{The }\textit{\textcolor{black}{disparity
filter }}\textcolor{black}{of \cite{serrano_extracting_2009} accomplishes
this by evaluating all edges incident on a given vertex in relation
to one another. The GloSS filter of \cite{radicchi_information_2011}
is a computationally involved method attempting to preserve the weight
distribution of edges.} Here we propose two new measures of significance
based on a different null model. The first, which we dub ``Marginal
Likelihood Filter'' (MLF) is a local significance measure computed
independently for each edge from the marginal probability distribution
of each edge weight, reducing pruning to a sorting problem. We judge
the significance of an edge in relation to the properties of both
of its end vertices. According to our null model, the higher the degrees
of two arbitrary vertices, the more likely they are to be connected
to one another by chance. Therefore, the higher the degrees of an
edge's incident vertices, the larger its weight must be for it to
be considered significant. The second, called the ``Global Likelihood
Filter'' (GLF) is a global measure computed for each possible subgraph
of a given size ``as a whole'', thus taking into account the correlation
among edges. Pruning then consists of finding the subgraph with the
highest significance. \textcolor{black}{While arguably the more principled
approach due to its global nature, GLF requires the use of Monte Carlo
methods and can thus become prohibitively expensive for large graphs.
MLF, on the other hand, provides a fast (almost linear in the number
of edges) method easily scalable to very large graphs.}

In the following sections we will define the null model and derive
from it the marginal likelihood edge filter for undirected as well
as directed weighted networks. Then we show how, from the same null
model, an ensemble approach to pruning can be developed and we describe
the resulting filter (GLF). We apply the methodology to the network
of air traffic volume between US airports in 2012, and demonstrate
how the filtered subgraphs differ in important topological measures
from those obtained from simple weight thresholding at a comparable
level.

\section{Marginal likelihood filter}

The null model defines a ``random'' ensemble of graphs resembling
the realized graph. We must therefore select some attributes of our
graph and demand that the random ensemble possess those attributes.
We propose a null model that preserves the total weight of the realized
graph and its degree sequence \textit{on average.} Here, by the degree
of a vertex we mean the sum of the weights of all its incident edges---also
known as the node's \textit{strength}---and we assume all weights
to be positive integers. Further, we conceive of a weighted edge as
multiple edges of unit weight.

For a weighted undirected graph then, our null model assumes that
the unit edges of the graph are assigned to a pair of vertices, one
at a time, and independently of one another. For each edge, the two
end points are chosen independently at random with probabilities proportional
to the degrees. That is, a vertex with a higher realized degree is
proportionally more likely to be assigned to an edge than a vertex
of lower degree. This leads to the same pair-wise connection probability
predicted by the \textit{configuration model }\cite{newman_structure_2003}\textit{
}in the \textit{sparse }or \textit{classical }limit \cite{park_statistical_2004}\textit{.
}Intuitively, vertices $i,j,\cdots$ in this model behave like chemical
reactants in a solution with concentrations $k_{i},k_{j},\dots,$
whose pairwise reaction rates are proportional to both reactant concentrations.
Given this null model, for any arbitrary pair of vertices $i$ and
$j$ with degrees $k_{i}$ and $k_{j}$, we can compute the probability
mass function of the weight of the edge connecting them.

Suppose the graph possesses a total of $T$ edges (recall that we
count a weighted edge as multiple edges of unit weight). \textcolor{black}{Throughout,
we will assume that $T\gg1.$ }Each \textcolor{black}{unit edge} must
choose two incident vertices at random, with probabilities proportional
to vertex degrees. The probability that $m$ out of the $T$ edges
will choose nodes $i$ and $j$ as their end points is given by the
binomial distribution $B(T,p).$ In short, the null model is defined
by the following distribution for the weight $\sigma_{ij}$ of the
$(i,j)$ pair: 
\begin{eqnarray}
\Pr\left[\sigma_{ij}=m\,|\,k_{i},k_{j},T\right] & = & {T \choose m}p^{m}(1-p)^{T-m}\\
\mbox{where}\,\,\,p=\frac{k_{i}k_{j}}{2T^{2}}, &  & T=\frac{1}{2}\sum_{i}k_{i}\label{eq:pij-main}
\end{eqnarray}
One can verify that the expected value of the degree of node $i$
is $\sum_{j}k_{i}k_{j}/(2T)=k_{i}.$ Thus, the ensemble defined by
the null model preserves the degree sequence \textit{on average. }We
note that depending on the value of $pT,$ for large $T$ this distribution
can tend to Poisson or normal distribution. With this distribution
at hand, we can proceed to compute a \textit{p}-value for the realized
value of the edge weight connecting $i,j.$ Denote the realized weight
of the $(i,j)$ edge by $w_{ij}.$ Then, we can define the \textit{p}-value
as 
\begin{equation}
s_{ij}(w_{ij})=\sum_{m\geq w_{ij}}\Pr\left[\sigma_{ij}=m\,|\,k_{i},k_{j},T\right].\label{eq:pvalue}
\end{equation}
This definition corresponds to a so-called \textit{one-tailed test}
where higher weights are considered more extreme regardless of the
expected value of the null distribution. Once we have computed the
\textit{p}-value for all edges, we can proceed to filter out any edge
with \textit{p}-value $s_{ij}(w_{ij})<\alpha$ for any threshold $\alpha$
of our choosing. This will retain the edges least likely to have occurred
purely by ``chance'' according to the marginal distributions resulting
from the null model. Numerical evaluation of the \textit{p}-value
from the binomial probability distribution will pose challenges due
to the large factorials involved. For large $T,$ one can use asymptotic
approximations of the binomial distribution instead (Poisson for $pT=O(1)$
and normal for $pT$$\gg1$). Some standard statistics packages include
implementations of the so-called \textit{binomial test} which computes
precisely the \textit{p}-value in question. We use the implementation
in Python's statsmodels package. 
\begin{figure*}
\medskip{}

\begin{raggedright}
\subfloat[\label{fig:subplot1}]{\includegraphics[height=5cm]{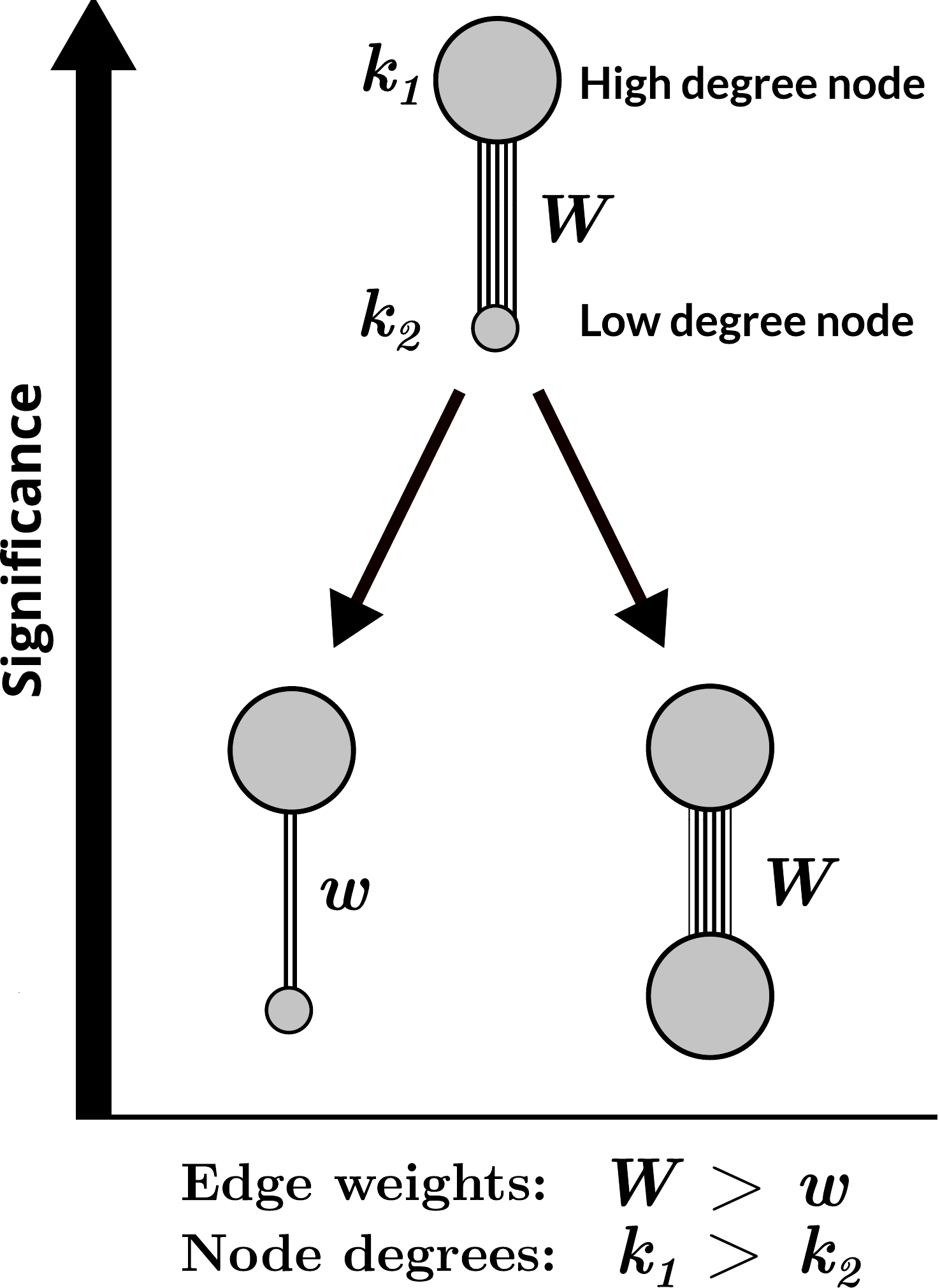}

}\quad{}\subfloat[\label{fig:subplot2}]{\begin{raggedright}
\includegraphics[height=5cm]{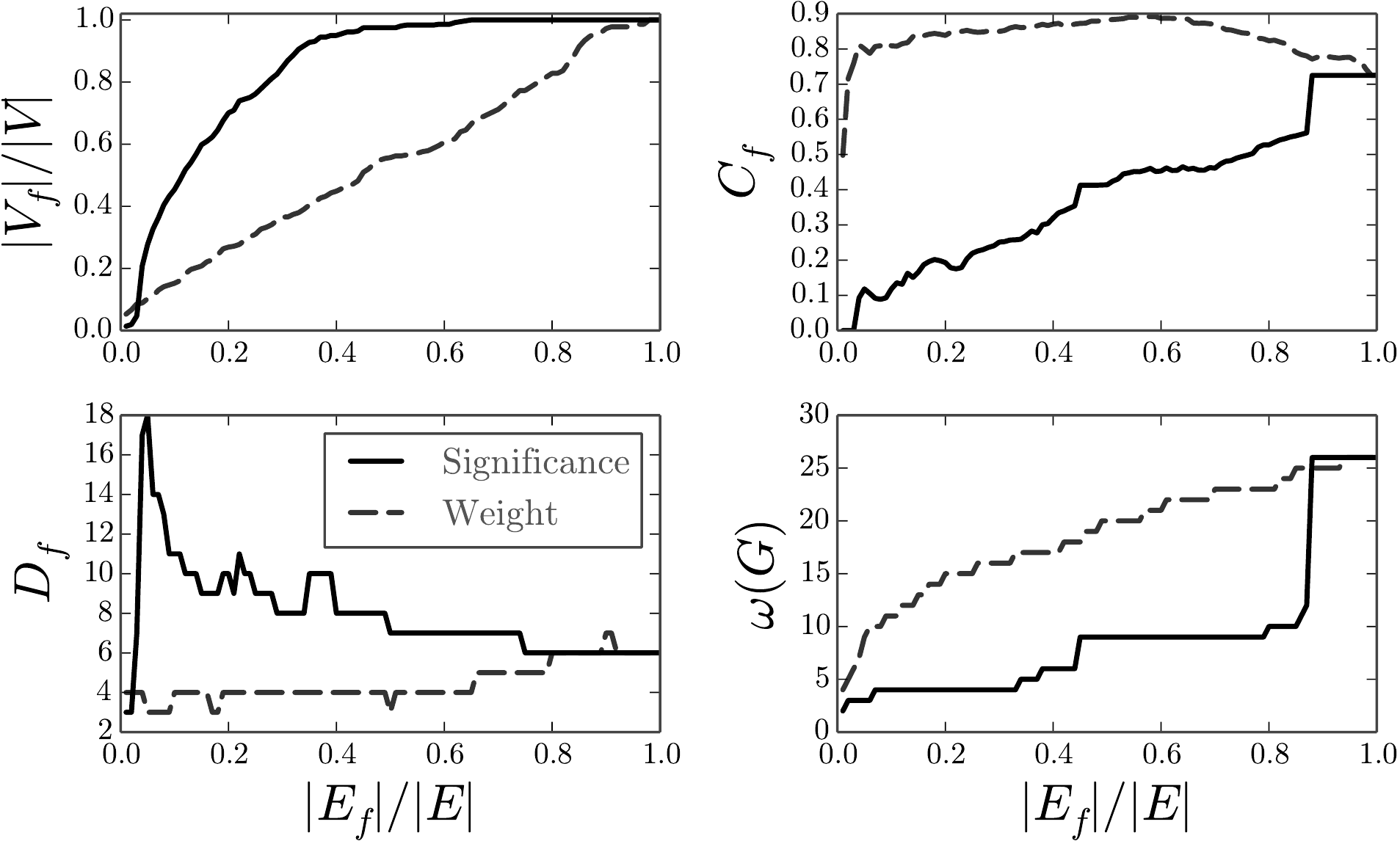}
\par\end{raggedright}

}\qquad{}\subfloat[]{\includegraphics[height=5cm]{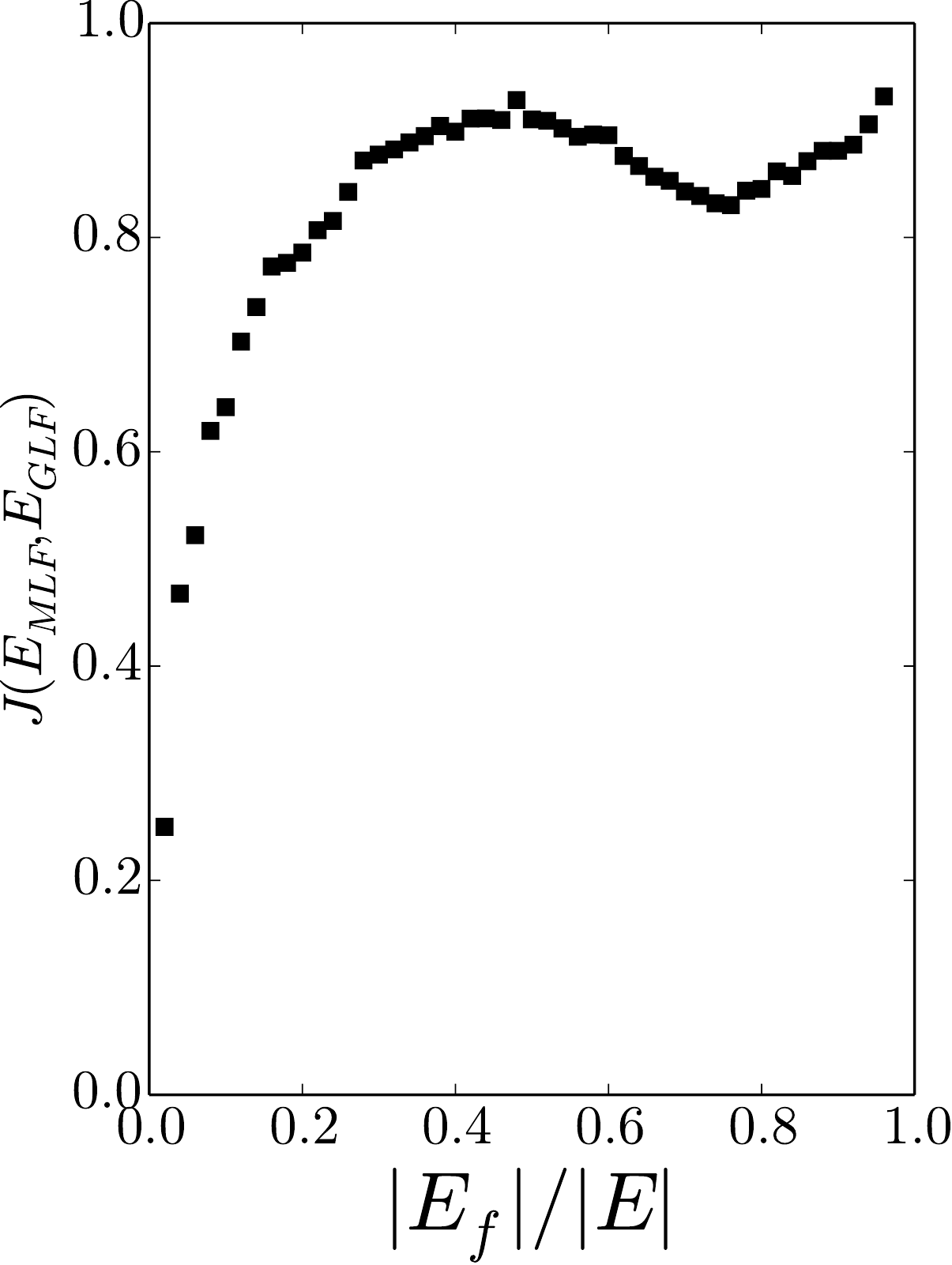}

}
\par\end{raggedright}

\caption{{\small{}(a) Qualitative schematic of the partial order defined by
the MLF filter: }\textcolor{black}{\small{}three pairs of nodes connected
by edges of varying weights. The size of a node represents its (weighted)
degree and the thickness of the edge represents its integer weight.}{\small{}
The top case has a higher significance than either of the bottom cases:
}\textcolor{black}{with the node degrees fixed, a higher edge weight
$W$ results in higher significance. With the weight $W$ fixed, lowering
either end-node's degree results in higher significance.}{\small{}
(b) Four graph measures computed for the US air traffic network (2012)
filtered at different levels using the Marginal Likelihood Filter
(solid) and weight thresholding (dashed). The $x$ axis is the proportion
of edges retained by the filtering. Clockwise from top left: 1- Proportion
of nodes in the giant component. 2- Clustering coefficient for the
giant component. 3- Diameter of the graph. 4- Clique number of the
graph. (c) The Jaccard similarity between the set of ``on edges''
produced by the MLF and GLF filters. \label{fig:figure-1}}}
\end{figure*}

We can generalize this formalism to the case of weighted directed
graphs. Here, the graph is characterized by two degree sequences:
the in-degree sequence, and the out-degree sequence. For a directed
edge between vertices $i,j$, the realized state consists of 
\begin{align}
 & \sigma_{ij}\,\,\mbox{weight of the directed edge }(i,j)\\
 & k_{i}^{out}\mbox{ out-degree of node }i\\
 & k_{j}^{in}\mbox{ \,\ in-degree of node \ensuremath{j}}
\end{align}
Again, we assume as the null model, that each of the $T$ directed
edges must choose a source vertex and a target vertex independently
at random, such that both the in-degree distribution and the out-degree
distribution reflect the realized values on average. Thus, the source
and target vertices must be chosen with probability proportional to
the nodes' out and in-degrees respectively. The weights will be distributed
binomially: 
\begin{eqnarray}
\Pr\left[\sigma_{ij}=m\,|\,k_{i}^{out},k_{j}^{in},T\right] & = & {T \choose m}p_{ij}^{m}(1-p_{ij})^{T-m}\\
\mbox{where}\,\,\,p_{ij}=\frac{k_{i}^{out}k_{j}^{in}}{T^{2}}, &  & T=\sum_{i}k_{i}^{out}
\end{eqnarray}
The \textit{p}-value is defined just as in $\left(\ref{eq:pvalue}\right),$
\textcolor{black}{replacing $k_{i}$ with $k_{i}^{out}$ and $k_{j}$
with $k_{j}^{in}.$}

\subsection{\textcolor{black}{Excluding self-edges}}

\textcolor{black}{The connectivity probability $\left(\ref{eq:pij-main}\right)$
used in our null model corresponds to the configuration model (more
specifically, its }\textit{\textcolor{black}{sparse limit}}\textcolor{black}{{}
\cite{park_statistical_2004}) in which self-edges are not precluded.
In most applications, whether the observed network contains self-edges
or not, this poses no problem in principle since randomization necessarily
involves }\textit{\textcolor{black}{forgetting}}\textcolor{black}{{}
certain properties of the observed instance. Furthermore, the sparse
limit of the configuration model is a good approximation of the special
loopless case when no vertex is dominant in terms of weighted degree.
However, if the preclusion of loops is fundamental to the underlying
dynamics or structure of the graph, one can easily modify the null
model to account for this fact. We find an alternative expression
for the connectivity probability $p_{ij}$ by seeking solutions of
the form 
\begin{equation}
p_{ij}=\begin{cases}
f(k_{i})f(k_{j}) & i\neq j\\
0 & i=j
\end{cases}
\end{equation}
 such that the degree sequence is preserved on average:
\begin{equation}
k_{i}=T\sum_{j\neq i}p_{ij}=Tf(k_{i})\left(\sum_{j}f(k_{j})-f(k_{i})\right).\label{eq:general-connectivity}
\end{equation}
The difference here is that $i$ is excluded from the sum. Note that
this equation automatically satisfies the loopless normalization condition
$\sum_{i<j}p_{ij}=1$ as well. To first order in $f(i)/\sum f(i),$
the solution is $f(i)=k_{i}/T\sqrt{2}$ and we recover $\left(\ref{eq:pij-main}\right).$
One can continue solving for higher order terms to compute the loopless
solution to arbitrary precision. Here we demonstrate the solution
of the second order term. Writing $f(k_{i})=f_{i}+\delta_{i}+O\left(\left(k_{i}/T\right)^{3}\right)$
where $f_{i}=k_{i}/T\sqrt{2}$ is the first order solution, and solving
for $\delta_{i}$ to second order, we find
\begin{equation}
\delta_{i}=\frac{f_{i}^{2}-cf_{i}}{\sqrt{2}}\,\,\,\mbox{where}\,\,\,c=\sum_{j}\delta_{j}.\label{eq:second-order}
\end{equation}
We need only solve for $c$ by summing over all $\delta_{i},$ which
yields
\begin{equation}
c=\frac{1}{\left(1+\sqrt{2}\right)}\sum_{i}f_{i}^{2}.\label{eq:second-order-C}
\end{equation}
We will continue to refer to the simplified model defined by $\left(\ref{eq:pij-main}\right)$
as the MLF, and to the modified versions defined by (\ref{eq:general-connectivity}),
(\ref{eq:second-order}) and (\ref{eq:second-order-C}) as the }\textit{\textcolor{black}{second
order loopless MLF.}}

\section{Global likelihood filter}

\textcolor{black}{The Marginal Likelihood Filter described above}
assigns a \textit{p}-value independently to each edge allowing us
to define the pruned graph simply as the subgraph consisting of the
top most significant edges. This results in a fast algorithm running
in $O(\left|E\right|\log\left|E\right|)$ time \textcolor{black}{where
$|E|$ is the size of the edge set of the graph.} However, this comes
at a cost: it is assumed that the statistical significance of an edge
is independent of the presence of other edges. In this section we
develop an ensemble approach to pruning that avoids this assumption.
In order to motivate this approach, we begin by reviewing the \textit{exponential
random graph model.}

The so-called exponential random graph model (ERGM) has been used
for decades---especially in the social sciences---to model unweighted
empirical networks and study the relationship between different graph
properties where simpler regression models fail due to the correlated
nature of the data. The model consists of a probability measure on
the set of all possible graphs with $n$ vertices given by 
\begin{equation}
P(G)\sim e^{\theta_{1}x_{1}(G)+\theta_{2}x_{2}(G)+\cdots\theta_{m}x_{m}(G)}
\end{equation}
where $x_{i}(G)$ is some graph property such as the total number
of edges or the degree of a specific node, etc., and $\theta_{i}$
is a parameter that must be adjusted such that $\left\langle x_{i}(G)\right\rangle $,
the ensemble average of property $x_{i}$, matches the observed value.
It was shown \cite{park_statistical_2004} that this probability measure
can be derived as the \textit{Boltzmann (or Gibbs) }distribution for
a canonical ensemble of graphs. In other words, this is the probability
measure that maximizes the entropy while keeping each graph property
$x_{i}$ equal to a fixed value \textit{on average, }just as the familiar
Boltzmann distribution of the canonical ensemble in statistical mechanics
yields a maximum entropy ensemble with a given average energy. The
parameter $\theta_{i}$ then acts as an inverse temperature whose
value adjusts $\left\langle x_{i}(G)\right\rangle .$ Given one such
model, one can compute other graph properties of interest as derivatives
of the partition function. 

\begin{figure*}
\subfloat[]{

\includegraphics[width=0.4\textwidth]{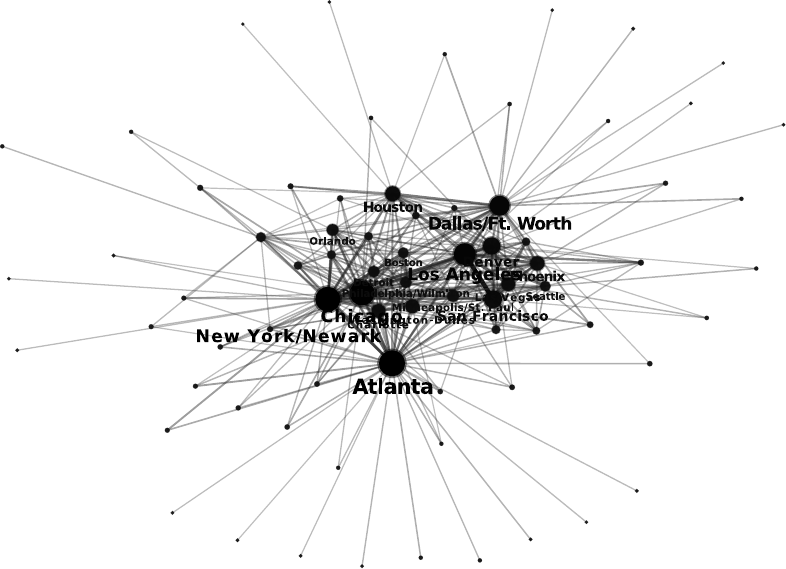}}\subfloat[]{\includegraphics[width=0.6\textwidth]{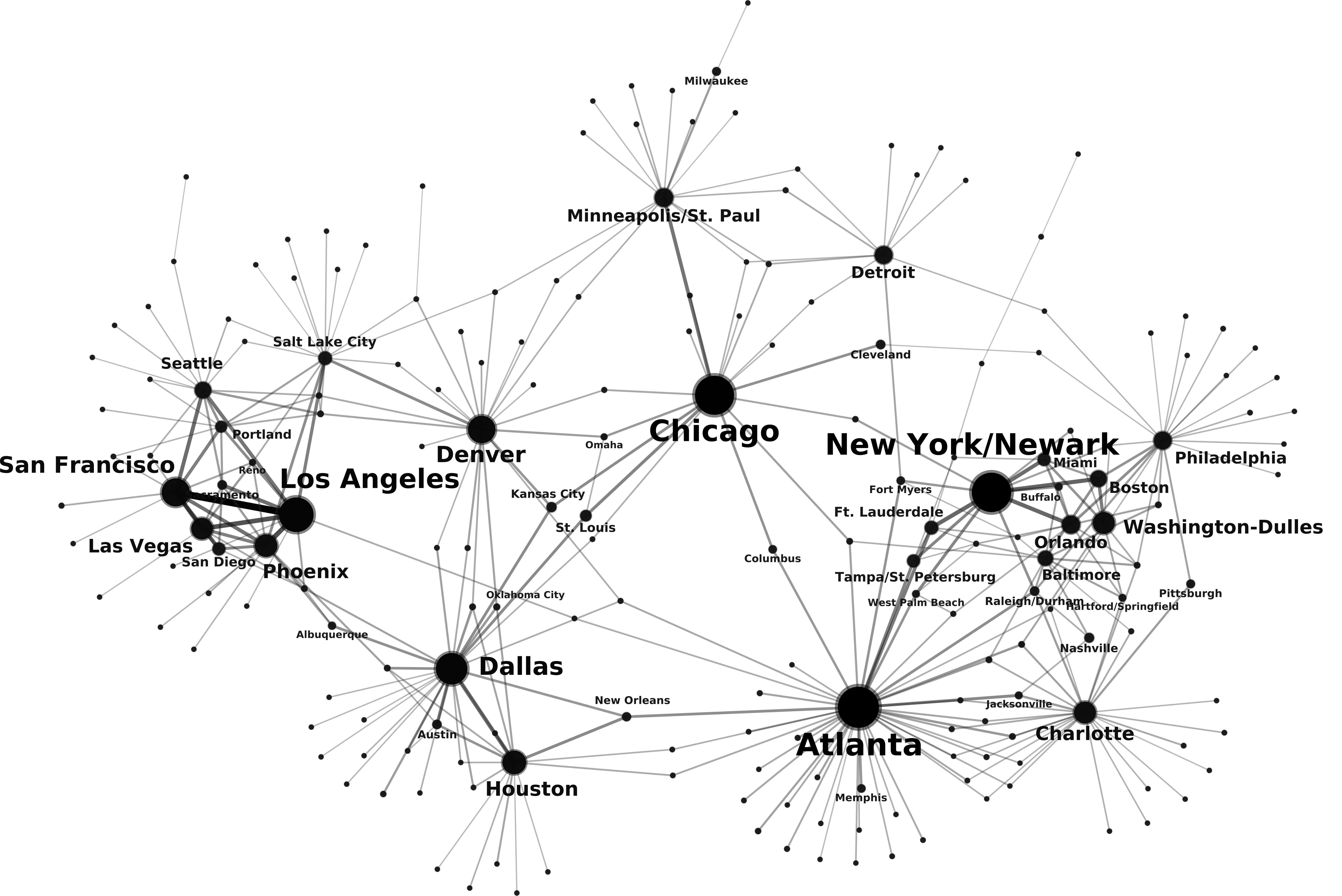}

}\caption{{\small{}Visualizations of the US airport transportation network (2012)
pruned using (a) weight thresholding, and (b) the Marginal Likelihood
Filter. In each case, the top 15\% of the edges with the respective
edge attribute are retained. Both plots are rendered using the same
standard Fruchterman-Reingold layout algorithm with identical parameters.\label{fig:Visualization-of-the}}}

\end{figure*}

More recently, the ERGM was generalized to weighted or multi-edge
graphs \cite{sagarra_configuration_2014,sagarra_statistical_2013}
where a weighted edge/multi-edge represents the number of ``events''
associated with a pair of nodes. Examples include transportation traffic
volume networks where the weight of the edge between two cities counts
the number of ``travel events'' between the two cities over a given
time period. Here, a pair of nodes can be viewed as a physical state
that can be occupied by an arbitrary number of particles, and the
weight of their incident edge corresponds to the occupation number
of the state. The crucial difference here is the \textit{distinguishability
}of the events as pointed out by \cite{sagarra_statistical_2013}.
Unlike physical particles, macroscopic events represented by multi-edge
graphs are distinguishable, and thus, the statistics resemble a Boltzmann
gas rather than a Bose-Einstein gas \cite{huang_introduction_2001}.
In concrete terms, this manifests as a multiplicity number associated
to each weighted graph configuration that must be taken into account
in the partition function.

Let us now define the GLF filtering procedure. Suppose we have an
observed graph $G_{0}$ with $n$ vertices and edge weights $w_{ij}\in\{0,1,2,\cdots\}$
for $i,j=1,2,\cdots,n$ so that the graph possesses $m$ weighted
edges (edges with positive weight). As the null model, we consider
the maximum entropy ensemble $\mathscr{G}$ of (integer) weighted
graphs with $n$ vertices such that the degree sequence is equal to
that of the observed graph on average. This automatically ensures
that the total edge strength of the graph is also equal to that of
the observed graph on average. Thus, following \cite{park_statistical_2004}
and \cite{sagarra_statistical_2013}, we obtain the grand canonical
ensemble defined by the Boltzmann distribution 
\begin{equation}
P(G)=\frac{1}{Z}g\left[\left\{ \sigma_{ij}\right\} \right]\exp\left[-\sum_{i<j}(\theta_{i}+\theta_{j})\sigma_{ij}\right]\,\,\,\forall G\in\mathscr{G}\label{eq:ensemble distribution}
\end{equation}
where $\sigma_{ij}\in\{0,1,2,\cdots\}$ is the weight of the $(i,j)$
edge in $G$, $\theta_{i}$ is the inverse temperature determining
$\left\langle k_{i}\right\rangle $, and 
\begin{equation}
g\left[\left\{ \sigma_{ij}\right\} \right]=\frac{\left(\sum_{i<j}\sigma_{ij}\right)!}{\prod_{i<j}\sigma_{ij}!}
\end{equation}

is the multiplicity of the configuration $\left\{ \sigma_{ij}\right\} $
resulting from the distinguishability of the events. Note that using
the partition function $Z,$ we can compute the parameters $\theta_{i}$
such that the ensemble's expected value of the total weight and the
degree sequence match those of the observed graph, namely $T=\frac{1}{2}\sum_{ij}w_{ij}$
and $\left\{ k_{i}\right\} $. In the thermodynamic limit $T\gg1,$
\textcolor{black}{up to an additive constant}, the log-likelihood
is given by 
\begin{equation}
\log P(G)=\log\left(\overline{N}!\right)+\sum_{i<j}\left[\sigma_{ij}\log p_{ij}-\log\left(\sigma_{ij}!\right)\right]\label{eq:logP}
\end{equation}
where 
\begin{equation}
p_{ij}=\frac{k_{i}k_{j}}{2T^{2}},\,\,\,\overline{N}=\sum_{i<j}\sigma_{ij}.\label{eq:pij}
\end{equation}
\begin{figure*}
\centering{}\subfloat[]{\includegraphics[width=0.4\textwidth]{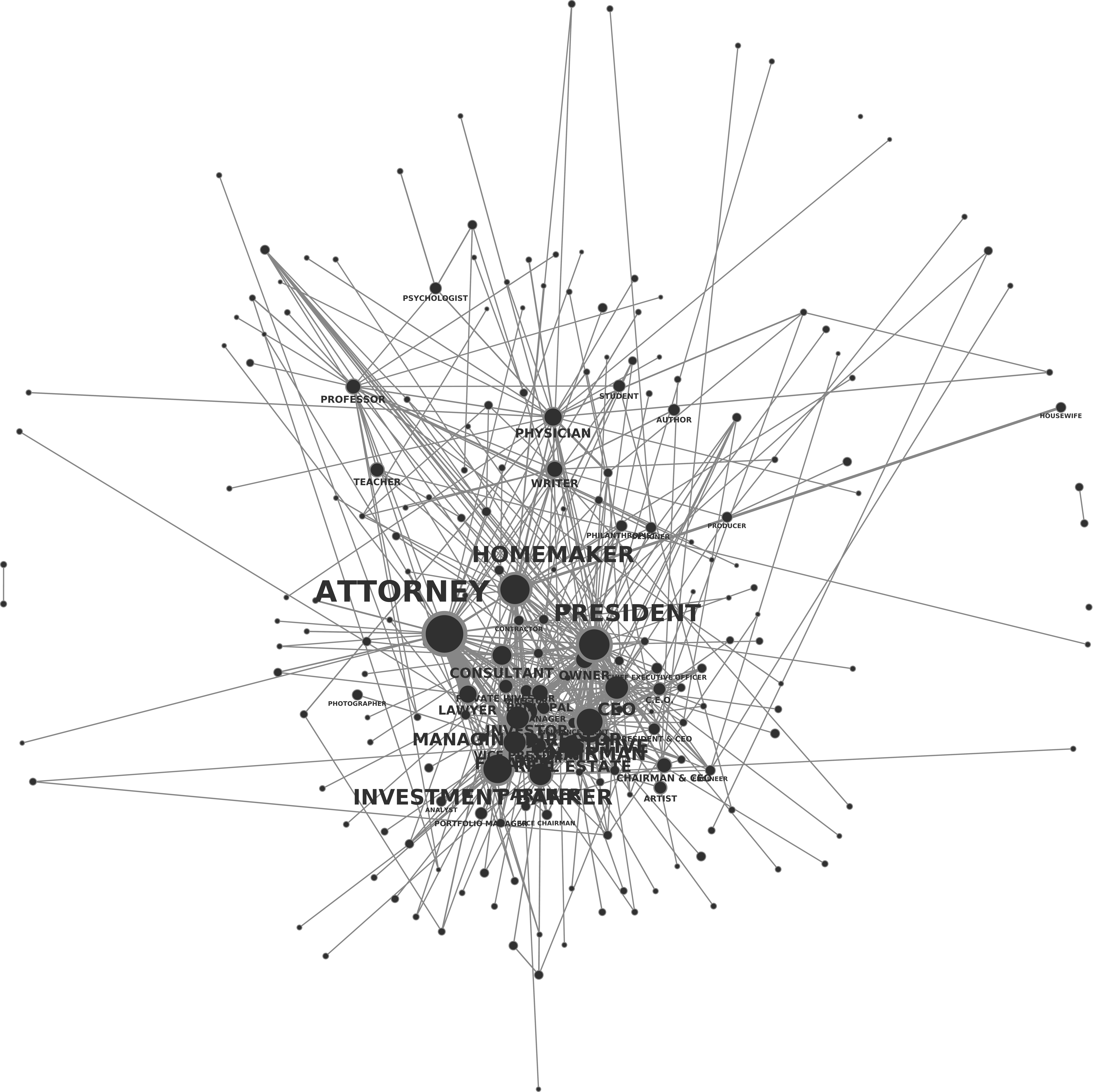}}~~~\subfloat[]{\includegraphics[width=0.6\textwidth]{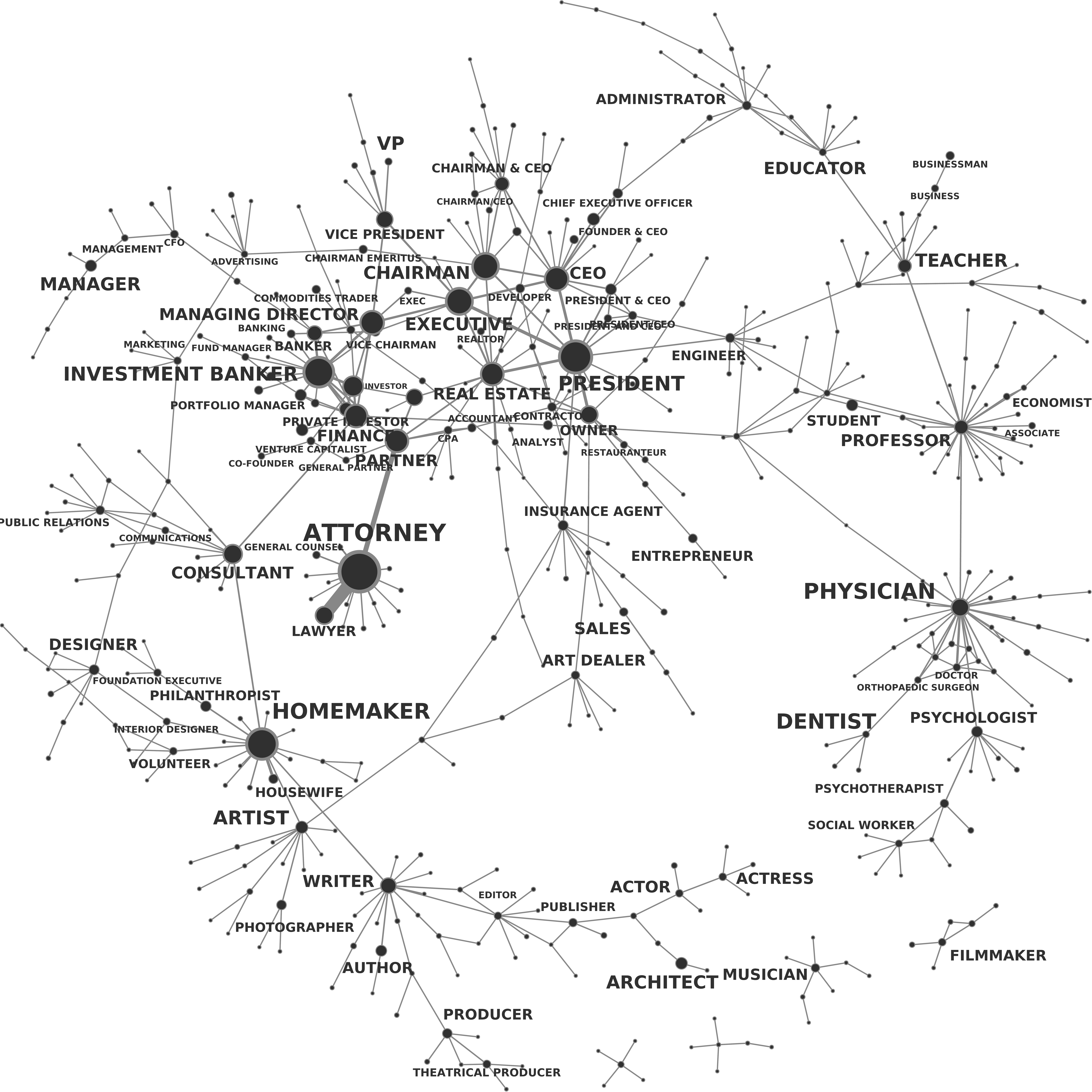}

}\caption{\textcolor{black}{\small{}Visualizations of the network of interdependent
occupations in the state of New York, }{\small{}pruned using (a) weight
thresholding, and (b) the Marginal Likelihood Filter.}\textcolor{black}{\small{}
Both are truncated at the 3\% level and plotted using a standard Fruchterman-Reingold
layout algorithm with identical parameters.}{\small{} \label{fig:occupations}}}
\end{figure*}
To evaluate the large factorials involved, one can use the highly
accurate Stirling approximation. For the details of the computations
leading to $\left(\ref{eq:logP},\ref{eq:pij}\right)$, see Appendix
A.

We posit that the most significant subgraph with $m'<m$ edges is
the \textit{minimum likelihood }subgraph among the set of all subgraphs
of $G_{0}$ possessing $m'$ non-zero weights, according to the null
distribution $\left(\ref{eq:ensemble distribution}\right)$. This
is the subgraph least likely to have been generated purely by chance
given the null distribution. Note that unlike the Marginal Likelihood
Filter discussed in the previous section, here we cannot assign independent
scores to edges and select the top $m'.$ The term $\log(\overline{N}!)$
combines the effect of all existing edges in a realization inseparably.
Therefore, solving this minimum likelihood problem requires a standard
Monte Carlo simulation. For instance, a Metropolis algorithm can be
used where initially a random set of $m'<m$ edges from $G_{0}$ are
``turned on'', and then at each step, one of the ``on'' edges
is chosen at random and replaced with a randomly selected ``off''
edge. The change is accepted if it decreases $\log P(G)$ and rejected
with probability $1-\exp(-\Delta\log P(G))$ otherwise, all the while
keeping track of the minimum likelihood set found thus far. The change
in log-likelihood is easy to compute since the second term in $\left(\ref{eq:logP}\right)$
is a sum over all ``on edges'', and $\Delta\log\left(\overline{N}!\right)$
can also be computed easily using the Stirling approximation.

\section{Application to real world networks}

In this section, we apply weight thresholding and the Marginal Likelihood
Filter to two networks. The first is the network of US air traffic
in 2012 (data courtesy of Alessandro Vespignani, following the work
in \cite{barrat_architecture_2004}.) In this network each node is
a US city and an edge weight represents the air traffic volume between
airport(s) in one city and another, aggregated over the year 2012.
The network is symmetrized and undirected.

Fig. \ref{fig:figure-1}(b) summarizes four graph measures computed
for this network truncated at different levels, both using the MLF
and using weight thresholding. The GLF filter yields results similar
to the MLF. The $x$ axis is the percentage of the total edges retained
in the truncated version. The four measures are the following: 1.
the size (number of nodes) remaining in the giant component $(\mbox{\ensuremath{|V_{f}|/|V|}}$)
2. the averaged local clustering for the giant component $C_{f}$
\cite{newman_structure_2003}. 3. the diameter of the graph $D_{f}$
4. the \textit{clique number }of the graph $\omega(G)$ \textcolor{black}{which
measures the size of the largest complete subgraph, or clique, found
within the graph} \cite{west_introduction_2001}. We observe that
at the same level of truncation, the significance filter leads to
a much larger giant component. Roughly at the 50\% level, almost all
nodes are already in the giant component. \textcolor{black}{When pruning
highly connected graphs, retaining a large giant component is naturally
desirable since decomposing the graph into a large number of small
components results in the loss of all connectivity information (e.g.
network distance) between most pairs of nodes.} \textcolor{black}{Even
if one hopes to accentuate the community structure of the graph via
pruning, it is still preferable for the communities to remain connected
to the giant component via weak links than to be completely cut off
as distinct connected components.} The clustering coefficient for
the weight-threshold truncations remains roughly the same for all
thresholds, whereas the significance filter produces considerably
lower clusterings at severe truncations, suggesting that the truncated
graph is rather sparse. The diameter (longest shortest path) of the
truncated graphs are also significantly different between the two
filters, with the significance filter yielding rather large diameters
at severe truncations, suggesting a sharper departure from a fully
connected graph. Finally, we observe a significant difference between
the clique numbers of the graphs according to the two filters. For
the weight filter, $\omega$ increases steadily as more and more edges
are included, whereas for the significance filter, it remains at a
more or less constant and low value until about the 90\% threshold
at which point a sharp increase brings it to the level of the untruncated
network. This reinforces the finding on the clustering number suggesting
that the significance threshold produces graphs with lower local densities.

Fig. \ref{fig:Visualization-of-the} compares the US airport transportation
network truncated using the MLF and weight thresholding. (Again, the
GLF produces results very similar to the MLF.) In both cases 15\%
of the edges are retained and the plots are rendered using a generic
force-directed layout algorithm (Fruchterman-Reingold) with identical
parameters. While the weight thresholded graph still appears as a
``hairball'' graph, the significance-filtered graph naturally unfolds
into what resembles the actual geographical distribution of the nodes
almost perfectly. This particular effect is in part due to the removal
of long-range high-volume edges that are nevertheless assigned a low
significance due to the high strength of their incident vertices.
For instance, the edges (Los Angeles, New York City) and (Chicago,
San Francisco) are absent from this truncation despite their large
weight. Our filter is thus prioritizing local connections over long-range
connections indicating the higher importance of these links with respect
to the overall traffic volume of their two end points. \textcolor{black}{This
is of course specific to this network, but demonstrates that }\textit{\textcolor{black}{some
}}\textcolor{black}{underlying property of the network is revealed
by pruning.} Finally, Fig. \ref{fig:figure-1}(c) compares the edge
sets selected by the MLF and GLF filters ($E_{MLF},E_{GLF})$ at different
truncation levels. The $y$ axis is the \textit{Jaccard similarity
}between the two edge sets defined as 
\begin{equation}
J(A,B)\equiv\frac{\left|A\cap B\right|}{\left|A\cup B\right|}.
\end{equation}
We observe that the two filters show a high level of similarity---over
80\% for truncation thresholds as low as 20\%. The disparity is of
course to be expected, since in the MLF scheme, the edge set at a
lower threshold is necessarily a subset of the edge set at a higher
threshold whereas in the GLF scheme, it is possible that an edge which
was absent at a higher threshold will be present at a lower threshold
(more severe truncation).

\textcolor{black}{We also applied the second order loopless MLF to
the air traffic network, and found that the pruned graph is virtually
indistinguishable from that obtained from the first order (standard)
MLF. To be specific, the two pruned graphs differed on a handful of
edges. Most notably, the Chicago-Minneapolis and Columbus-Atlanta
edges were absent from the result of the loopless MLF, but barely
changed the graph layout, suggesting that the standard MLF can serve
as a good approximation to the loopless case as well.}

\textcolor{black}{Next we applied the MLF to the network of interdependent
occupations in the state of New York derived from the database of
campaign contributions published by the Federal Election Commission
(FEC). This database contains every monetary contribution over \$200
by an individual to a federal US election campaign since 1979. Among
other information, each record contains the occupation and employer
of the donor at the time of the transaction. The author has compiled
a ``disambiguated'' version of this database (to be published separately
in the near future), meaning one in which all transactions from the
same individual are linked. This database consists of roughly 24,000,000
records from around 6,000,000 individual. Using this disambiguated
database one can produce a }\textit{\textcolor{black}{co-occurrence
network }}\textcolor{black}{of occupations where each node is a distinct
occupation label and two labels are linked if both appear in an individual's
history. Thus, an edge can indicate either semantic equivalence (``Doctor''
and ``Physician'') or a plausible transition (``Postdoc'' to ``Professor'').
The weight of an edge counts the number of times the two end-nodes
co-occurred in histories of individuals. For instance, ``Lawyer''
and ``Attorney'' are linked with a rather large weight. Similarly,
``President'' and ``CEO'' are strongly linked. Due to the large
size of the database however, many apparently unrelated occupations
are also linked, albeit with lower weights, e.g., a ``Doctor'' who
at some other time identifies as a ``Writer''. We can now ask if
we can uncover the structure of interrelated occupations by pruning
the dense co-occurrence network. Figure \ref{fig:occupations} shows
this network pruned using weight thresholding and the MLF. In both
cases, the top 3\% of the edges are retained. As with the air transportation
network, weight thresholding does little to reveal the important structures
within the network while the MLF untangles the network into clearly
visible clusters: legal professions are connected through ``Partner''
to the large cluster of top management occupations, ``Homemaker''
is in a distinct cluster together with creative and philanthropic
occupations, and medical and academic occupations are each in their
own clusters.}

\section{Discussion}

In order to extract the most significant substructures in a complex
network, we have proposed a generative null model, and studied two
edge filters resulting from this model. Our significance measures
are derived from a null model that preserves the total edge strength
and the weighted degree sequence of the graph on average. Simply put,
this null model states that if everything were random, two arbitrary
vertices would be connected with probability proportional to both
their weighted degrees (strengths). In the first filter, the degree
of deviation from this null model in the observed network at the edge
level, expressed as a \textit{p-}value, defines the marginal significance
of an edge. When applied to real-world networks, this filter extracts
subgraphs that are significantly sparser (as measured by clustering,
clique number and shortest path length) than one would obtain from
simple weight thresholding at the same level, even though it yields
higher global connectivity as reflected by the size of the giant component.
In the second filter, the likelihood of the occurrence of each subgraph
\textit{as a whole }determines its global significance, and the most
significant subgraph corresponds to the \textit{minimum likelihood
}subgraph. Visual inspection of the US airport transportation network
filtered using our significance measures reveals how low-weight regional
links are prioritized over high-weight long-range links such that
the original ``hairball'' network unfolds into a rather flat graph
closely reflecting the actual geographical distribution of the nodes.

\textcolor{black}{Regarding the relationship between the two filters,
one more point merits discussion. First, equations $\left(\ref{eq:general-connectivity}\right)$
and $\left(\ref{eq:GLF-nonlinear-system}\right)$ are equivalent.
Furthermore, the Gibbs distribution found in $\left(\ref{eq:GLF-multinomial-dist}\right)$
is simply a multinomial distribution when the total edge weight is
constrained. It is a standard result that when the joint distribution
of multiple variables is multinomial, the marginal distribution of
each variable is simply binomial. We see then, that the edge weight
distribution used in MLF is simply the marginal distribution of the
Gibbs distribution used in GLF. }

\textcolor{black}{Let us also emphasize the fact that unlike in sparsification,
the goal in pruning is to reveal unknown structures which are obscured
by noise.} This is why the problem of pruning is not an approximation
problem as there are no objective measures of success. Therefore,
the merits of a pruning filter such as ours can only be judged by
the assumptions defining the null model. A central concern is the
balance between over and under-determination. The null model should
sufficiently reflect the essential features of the observed graph
through its defining constraints. However, one must avoid imposing
too many constraints or else the null model will be incapable of accounting
for the natural variations in the (unknown) ensemble from which the
observed graph is obtained. In this context, our filter can be viewed
as a middle ground between the disparity filter \cite{serrano_extracting_2009}
which defines the null model only based on a node's incident edges,
and the GloSS filter of \cite{radicchi_information_2011} which preserves
the global distribution of edge weights.

\textcolor{black}{Finally, we outline a number of potential future
directions. For highly skewed degree distributions, the asymptotic
expansions of different equations in $\left(\ref{eq:general-connectivity}\right)$
may need to be truncated at different orders depending on $k_{i}/T$
in order to produce balanced solutions. Another remaining question
is whether one can define an }\textit{\textcolor{black}{optimal }}\textcolor{black}{significance
threshold for pruning a given graph. Presumably, some combination
of the graph measures discussed in fig. \ref{fig:subplot2} as well
as other relevant measures can aid the practitioner in determining
the appropriate truncation level. However, whether or not a generically
applicable recipe can be defined remains to be seen.}

\section*{Acknowledgments}

The author wishes to thank Elaine Stranahan, Nima Dehmamy and Oleguer
Sagarra for fruitful discussions. This material is based upon work
supported in part by the National Science Foundation under grant No.
502019.

\appendix

\section{~}

In this appendix, we will compute the partition function for the ensemble
defined by $\left(\ref{eq:ensemble distribution}\right)$ and enforce
the constraints on the degree sequence by solving for the parameters
$\theta_{i},\,i=1,2,\cdots,n.$ The partition function is defined
by 
\begin{eqnarray}
Z & = & \sum_{\left\{ \sigma_{ij}\right\} }g\left[\left\{ \sigma_{ij}\right\} \right]\exp\left[-\sum_{i<j}(\theta_{i}+\theta_{j})\sigma_{ij}\right]\\
 & = & \sum_{N=0}^{\infty}\sum_{\substack{\left\{ \sigma_{ij}\right\} \\
\sum\sigma_{ij}=N
}
}N!\prod_{i<j}\dfrac{1}{\sigma_{ij}!}e^{-(\theta_{i}+\theta_{j})\sigma_{ij}}.
\end{eqnarray}
Using the Multinomial theorem, the inner sum simplifies: 
\begin{eqnarray}
Z & = & \sum_{N=0}^{\infty}\left(\sum_{i<j}e^{-(\theta_{i}+\theta_{j})}\right)^{N}\\
 & = & \frac{1}{1-\sum_{i<j}e^{-(\theta_{i}+\theta_{j})}}.
\end{eqnarray}
The expected values of the degrees can be computed as follows: 
\begin{align}
\left\langle \sum_{j\neq i}\sigma_{ij}\right\rangle =-\frac{\partial}{\partial\theta_{i}}\log Z
\end{align}
Setting these equal to the $k_{i}$ respectively and defining 
\begin{eqnarray}
x_{i} & \equiv & e^{-\theta_{i}}\,\,i=1,2,\cdots,n
\end{eqnarray}
after rearranging the terms we obtain

\begin{align}
k_{i} & =\frac{x_{i}\sum_{j\neq i}x_{j}}{1-\sum_{i<j}x_{i}x_{j}}\\
\sum k_{i} & =2T
\end{align}
Summing the first equation over $i$ and using the second equation
yields $\sum_{i<j}x_{i}x_{j}=T/(1+T).$ Thus we obtain a system of
$n$ nonlinear equations
\begin{equation}
\frac{k_{i}}{1+T}=x_{i}\left(\sum_{j=1}^{n}x_{j}-x_{i}\right)\,\,\,i=1,2,\cdots,n.\label{eq:GLF-nonlinear-system}
\end{equation}
\textcolor{black}{Note that this is identical to equation $\left(\ref{eq:general-connectivity}\right)$
arising in the loopless MLF.} Using the fact that $\sum_{i}k_{i}=2T\gg1,$
to first order in terms of $x_{i}/\left(\sum_{j}x_{j}\right)$ the
solution is
\begin{equation}
e^{-\theta_{i}}\equiv x_{i}\simeq\frac{k_{i}}{T\sqrt{2}}.
\end{equation}
Therefore, the probability distribution $\left(\ref{eq:ensemble distribution}\right)$
becomes 
\begin{equation}
P(G)=\frac{1}{Z}g\left[\left\{ \sigma_{ij}\right\} \right]\prod_{i<j}\left(\frac{k_{i}k_{j}}{2T^{2}}\right)^{\sigma_{ij}}\,\,\,\forall G\in\mathscr{G}\label{eq:GLF-multinomial-dist}
\end{equation}
and we obtain $\left(\ref{eq:logP}\right).$ 

\bibliographystyle{ieeetr}
\bibliography{GraphPruning,Networks}

\end{document}